\documentclass[submission,copyright,creativecommons]{eptcs}
\providecommand{\event}{PLACES 2022} 
\usepackage{underscore}           

\usepackage[utf8]{inputenc}

\usepackage[T1]{fontenc}
\usepackage{adjustbox}
\usepackage{amsmath}
\usepackage{amsfonts} 
\usepackage{float}  

\usepackage{stmaryrd}
\usepackage[draft]{fixme} 
\fxsetup{theme=color,mode=multiuser}
\FXRegisterAuthor{JL}{aJL}{JL}
\FXRegisterAuthor{TC}{aTC}{TC}
\FXRegisterAuthor{DL}{aDL}{DL}
\FXRegisterAuthor{HZ}{aHZ}{HZ}

\usepackage{mathpartir}
\usepackage{nicefrac}
\usepackage{listings}
\usepackage{pgfkeys}
\usepackage{keycommand}
\usepackage{amsthm,amsmath,amssymb}
\newtheorem{theorem}{Theorem}

\newtheorem{definition}{Definition}
\usepackage[left=30mm, right=30mm]{geometry} 
\usepackage{graphicx}   
\usepackage[inline]{enumitem}   
\usepackage{booktabs}   
\usepackage{tabularx}   
\usepackage{multirow}   
\usepackage{arydshln}   
\usepackage{subcaption} 
\usepackage{wrapfig}     

\usepackage[normalem]{ulem}
\usepackage[table]{xcolor}
\usepackage{lstcoq} 
\usepackage{hyperref}
\usepackage[capitalise,nameinlink,noabbrev]{cleveref}
\usepackage{tikz}
\usepackage[framemethod=TikZ]{mdframed}
\usepackage{cite}
\usepackage{ebproof} 
\usepackage{varwidth}
\DeclareMathAlphabet{\mathcal}{OMS}{cmsy}{m}{n}

\usetikzlibrary{tikzmark,positioning,patterns}
\tikzset{
  venn/.style={preaction={fill, #1},opacity=0.6,anchor=south}
}
\include{macros-balloon}
\usetikzlibrary{shadows.blur}
\usepackage{xspace}
\newcommand{\eg}{\textit{e.g.},\@\xspace}
\newcommand{\ie}{\textit{i.e.},\@\xspace}

\newcommand{\etal}{\textit{et al.}\@\xspace}

\newcommand{\miniparagraph}[1]{%
  \smallskip
  \noindent
  \textbf{#1.} }

\newcounter{RQcounter}

\newcommand{\RQref}[1]{\hyperref[rq:#1]{Claim~\ref*{rq:#1}}}

\newcommand{\Plang}{\mathcal{S}}

\newcommand{\Clang}{\mathcal{U}}
\newcommand{\BabyClang}{\mathcal{B}}
\newcommand{\pad}{\;\;}
\newcommand{\Space}[1]{\pad{#1}\pad}
\newcommand{\Grmeq}{{::=}}
\newcommand{\grmeq}{\Space\Grmeq}

\newcommand{\Grmor}{\mid}
\newcommand{\grmor}{\;\Grmor\;}
\newcommand{\eqdef}{\stackrel{\mathrm{def}}{=}}
\newcommand{\st}{\colon}

\DeclareMathOperator{\Bstep}{\downarrow}

\newcommand{\update}[2]{[{#1}\mapsto{#2}]}
\newcommand{\subs}[2]{[ #2 \coloneqq #1 ]}

\newcommand{\WriteAcc}[2]{{#1}:{#2}}

\newcommand{\FreeVar}[1]{\mathit{fv}(#1)}

\usepackage{mathtools}

\newcommand{\Nil}{[\;]}
\DeclareMathOperator{\Cons}{::}
\DeclareMathOperator{\Concat}{+\kern -0.4em+}

\newcommand{\TidDom}{\mathcal{T}}

\newcommand{\Tid}{i}
\newcommand{\Tidy}{j}
\newcommand{\Tidz}{k}

\newcommand{\NatDom}{\mathbb N}

\newcommand{\Nat}{\Tid}
\newcommand{\Naty}{\Tidy}
\newcommand{\Natz}{\Tidz}

\newcommand{\Nexp}{n}
\newcommand{\Nexpy}{m}

\newcommand{\Var}{x}
\newcommand{\Vary}{y}

\newcommand{\TidVar}{\kw[tid]{tid}}

\DeclareMathOperator{\Nbin}{\star}

\newcommand{\Bexp}{c}

\newcommand{\Bool}{\mathbb{B}}
\newcommand{\True}{\mathtt{true}}
\newcommand{\False}{\mathtt{false}}

\DeclareMathOperator{\Nrel}{\diamond}

\DeclareMathOperator{\Bbin}{\circ}

\newcommand{\Loc}{l}

\newcommand{\Mode}{o}

\newcommand{\Mread}{\kw[sync]{rd}}
\newcommand{\Mwrite}{\kw[sync]{wr}}

\pgfkeys{
  /acc/.is family,
  /acc/.cd,
  loc/.initial=\Loc, 
  loc/.get=\GetLoc, 
  loc/.store in=\GetLoc, 
  index/.initial=\Nexp,
  index/.get=\GetIndex,
  index/.store in=\GetIndex,
  mode/.initial=\Mode,
  mode/.get=\GetMode,
  mode/.store in=\GetMode
}
\newcommand{\Bwr}[2][\Nexp]{\mathtt{A}[{#1}] := {#2}}
\newcommand{\Brd}[3][\Var]{\kw{let} \; {#1} = \mathtt{A}[{#2}] \; \kw{in} \; {#3}}
\newcommand{\Wr}[1][\Nexp]{\Acc[index={#1},mode=\Mwrite]}
\newcommand{\Rd}[1][\Nexp]{\Acc[index={#1},mode=\Mread]}
\newcommand{\Acc}[1][]{%
  \pgfkeys{/acc/.cd,index=\Nat,#1}
  \GetMode\Index[\GetIndex]%
}

\newcommand{\Index}[1][\Nexp]{[{#1}]}

\newcommand{\Proto}{p}

\newcommand{\Psync}{\kw[sync]{sync}}

\newcommand{\By}[2][\Tid]{{#1}\colon\!{#2}}

\pgfkeys{
  /for/.is family,
  /for/.cd,
  var/.initial=\Var, 
  var/.get=\GetForVar, 
  var/.store in=\GetForVar, 
  lo/.initial=\Nexp,
  lo/.get=\GetForLo,
  lo/.store in=\GetForLo,
  hi/.initial=\Nexpy,
  hi/.get=\GetForHi,
  hi/.store in=\GetForHi,
  body/.initial=\Proto,
  body/.get=\GetForBody,
  body/.store in=\GetForBody,
}
\newcommand{\KforS}{\kw[decl]{for^{\mathtt{S}}}}
\newcommand{\Pfor}[1][]{%
  \pgfkeys{/for/.cd,#1}
  \KforS\ \GetForVar \in \GetForLo..\GetForHi\ \{\GetForBody\}%
}

\newcommand{\KforU}{\kw[decl]{for^{\mathtt{U}}}}
\newcommand{\Cfor}[1][]{%
  \pgfkeys{/for/.cd,body=\Cexp,#1}
  \KforU\ \GetForVar \in \color{black} \GetForLo..\GetForHi\ \{\GetForBody\}%
}
\newcommand{\BabyKforU}{\kw[decl]{for^{\mathtt{U}}}}
\newcommand{\BabyCfor}[1][]{%
  \pgfkeys{/for/.cd,body=\Cexp,#1}
  \BabyKforU\ \GetForVar \in \color{black} \GetForLo..\GetForHi\ \{\GetForBody\}%
}

\newcommand{\Pskip}{\kw[decl]{skip}}

\newcommand{\Cif}[3][\Bexp]{\kw[decl]{if}\ {#1}\ \{{#2}\}\ \kw[decl]{else}\ \{{#3}\}}

\pgfkeys{
  /decl/.is family,
  /decl/.cd,
  var/.initial=\Var, 
  var/.get=\GetVar, 
  var/.store in=\GetVar, 
  lo/.initial=\Nexp,
  lo/.get=\GetDeclLo,
  lo/.store in=\GetDeclLo,
  hi/.initial=\Nexpy,
  hi/.get=\GetDeclHi,
  hi/.store in=\GetDeclHi,
  body/.initial=\Proto,
  body/.get=\GetDeclBody,
  body/.store in=\GetDeclBody,
}

\newcommand{\Hist}{P}
\newcommand{\Histy}{Q}

\DeclareMathOperator{\Seq}{\, ; \,}
\DeclareMathOperator{\InferTo}{ \; \blacktriangleright \;}

\newcommand{\Cexp}{u}
\newcommand{\Cexpx}{u_1}
\newcommand{\Cexpy}{u_2}

\newcommand{\BabyCexp}{b}
\newcommand{\BabyCexpx}{b_1}
\newcommand{\BabyCexpy}{b_2}

\newcommand{\AccVal}{\alpha}

\newcommand{\Mhist}{H}
\newcommand{\Mhisty}{I}

\newcommand{\Vset}{\mathcal V}
\newcommand{\RWset}{A}
\newcommand{\RWsety}{B}
\newcommand{\RWsetz}{C}
\newcommand{\Writeset}{ W}
\newcommand{\Writesetx}{ W_1}
\newcommand{\Writesety}{ W_2}

\newcommand{\Readset}{ R}
\newcommand{\Readsetx}{ R_1}
\newcommand{\Readsety}{ R_2}

\pgfkeys{
  /tred/.is family,
  /tred/.cd,
  hist/.initial=\Mhist, 
  hist/.get=\GetHist, 
  hist/.store in=\GetHist, 
  tid/.initial=\Tid,
  tid/.get=\GetTid,
  tid/.store in=\GetTid,
}
\newcommand{\Tred}[1][]{%
  \pgfkeys{/tred/.cd,#1}
  \;\Bstep_{\GetHist,\GetTid}\;
}

\newcommand{\TredIn}[2][\RWset]{{#1},{#2}}

\pgfkeys{
  /rw/.is family,
  /rw/.cd,
  r/.initial=\Readset, 
  r/.get=\GetRead, 
  r/.store in=\GetRead, 
  w/.initial=\Writeset,
  w/.get=\GetWrite,
  w/.store in=\GetWrite,
}
\newcommand{\RW}[1][]{%
  \pgfkeys{/rw/.cd,#1}
  (\GetRead,\GetWrite)
}

\newcommand{\PredIn}[2][\Mhist]{{#1},{#2}}
\newcommand{\Pred}{\Bstep}

\definecolor{purple}{rgb}{0.65, 0.12, 0.82}
\definecolor{olive}{rgb}{0,0.40,0}
\newcommand{\kw}[2][sync]{\mathsf{\color{#1}{#2}}}

\newcommand{\drf}{DRF\xspace}

\newcommand{\lang}{MAP\xspace}
\newcommand{\Lang}{MAP\xspace}
\newcommand{\langs}{MAPs\xspace}

\newcommand{\babylang}{BabyCUDA\xspace}
\newcommand{\Tool}{{\sf{Faial}}\xspace}

\def\fcmp{\mathbin{\raise 0.6ex\hbox{\oalign{\hfil$\scriptscriptstyle%
\mathrm{o}$\hfil\cr\hfil$\scriptscriptstyle\mathrm{9}$\hfil}}}}

\def\fcmp{\mathbin{\raise 0.6ex\hbox{\oalign{\hfil$\scriptscriptstyle\mathrm{o}$\hfil\cr\hfil$\scriptscriptstyle\mathrm{9}$\hfil}}}}

\newcommand{\Ntypes}[1][\Vset]{{#1} \vdash \,}
\newcommand{\Btypes}[1][\Vset]{{#1} \vdash \,}
\newcommand{\Uinfer}[2][\Vset]{{#1}\, \vdash \,{#2}}
\newcommand{\Sinfer}[2][\Vset]{{#1}\, \vdash \,{#2}}
\newcommand{\NUinfer}[2][\Vset]{{#1}\, \nvdash \,{#2}}

\newcommand{\Lastwrite}[2][\Nat]{\mathit{lastwrite}(#1,\,#2)}

\newenvironment{ruled}{
  \hrule
  \vspace{1ex}
}{
  \vspace{1ex}
  \hrule
}

\usepackage{color}
\usepackage{xcolor}
\definecolor{base00}{HTML}{868e96}
\definecolor{base01}{HTML}{586E75}
\definecolor{base02}{HTML}{073642}
\definecolor{base03}{HTML}{495057}
\definecolor{base3}{HTML}{FDF6E3}
\definecolor{base2}{HTML}{EEE8D5}
\definecolor{base1}{HTML}{93A1A1}
\definecolor{base0}{HTML}{F8F9FA}
\definecolor{solyellow}{HTML}{B58900}
\definecolor{solorange}{HTML}{CB4B16}
\definecolor{solred}{HTML}{DC322F}
\definecolor{solmagenta}{HTML}{D33682}
\definecolor{solviolet}{HTML}{6C71C4}
\definecolor{solblue}{HTML}{268BD2}
\definecolor{solcyan}{HTML}{2AA198}
\definecolor{solgreen}{HTML}{859900}

\definecolor{tid}{HTML}{586E75}
\definecolor{decl}{HTML}{084c61}
\definecolor{sync}{HTML}{177e89}

\definecolor{pblue}{rgb}{0.13,0.13,1}
\definecolor{pgreen}{rgb}{0,0.5,0}
\definecolor{pred}{rgb}{0.9,0,0}
\definecolor{pgrey}{rgb}{0.46,0.45,0.48}

\usepackage{listings}

\lstdefinelanguage{Map}{%
  showspaces=false,
  showtabs=false,
  breaklines=true,
  showstringspaces=false,
  breakatwhitespace=true,
  columns=flexible,
  moredelim=[il][\textcolor{pgrey}]{$$},
  keywords={sfor,for,while,let,if,else,int,foreach,var},
  keywords=[2]{\_\_syncthreads,sync,rd,wr,skip},
  keywords=[3]{tid},
  literate={?}{}1
    {.}{\texttt{.}}1
    {<}{\texttt{<}}1
    {+}{\texttt{+}}1
    {*}{\texttt{*}}1
    {=}{\texttt{=}}1
    {:}{\texttt{:}}1
    {inSymbol}{$\in$}1
    {forU}{$\KforU$}3
    {forS}{$\KforS$}3
    ,
  moredelim=[is][\uwave]{`}{`},
  moredelim=[is][\textcolor{pgrey}]{@}{@},
  escapeinside={^}{^},
}
\lstdefinestyle{map}{%
  commentstyle=\color{pgreen},
  keywordstyle=\bfseries\color{decl},
  keywordstyle=[2]\bfseries\color{sync},
  keywordstyle=[3]\bfseries\color{tid},
  stringstyle=\color{pred},
  basicstyle=\sffamily\scriptsize,
  numberstyle=\tiny,
}

\lstnewenvironment{code}[1][]
{\lstset{language=Map,style=map,xleftmargin=1.7ex,numbers=left,#1}}
{}

\lstdefinestyle{inline}{%
  basicstyle=\normalsize\sffamily,
  numbers=none,
  backgroundcolor=,
  columns=fullflexible,
}

\usepackage{float}

\makeatletter
\newcommand\Xfloatc@ruled[2]{{\@fs@cfont #1} #2\par}
\newcommand\fs@ruledx{%
  \def\@fs@cfont{}\let\@fs@capt\Xfloatc@ruled%
  \def\@fs@pre{}
  \def\@fs@post{\kern2pt\hrule\relax}%
  \def\@fs@mid{\kern2pt\hrule\kern2pt}%
  \let\@fs@iftopcapt\iftrue}
\makeatother

\floatstyle{ruledx}
\newfloat{listing}{tbp}{lop}[section]
\floatname{listing}{Listing}

\newenvironment{sproof}{%
  \proof}{\endproof}

\def\titlerunning{Provable GPU Data-Races in Static Race Detection}
\title{\titlerunning}
\date{}
\def\authorrunning{D.\ Liew, T.\ Cogumbreiro, \& J.\ Lange}
\author{Dennis Liew$^1$
\qquad Tiago Cogumbreiro$^1$
\qquad Julien Lange$^2$
\institute{$^1 $ University of Massachusetts Boston, Boston, USA
\qquad
$^2 $ Royal Holloway, University of London, Egham, UK}
\email{\textbf{zhenrong.liew001@umb.edu}, \textbf{tiago.cogumbreiro@umb.edu},
  \textbf{julien.lange@rhul.ac.uk}
}}

\begin{document}

\maketitle

\begin{abstract}
  We extend the theory behind the \Tool tool-chain, which can soundly prove
  that CUDA programs (aka, kernels) are data-race free using
  specialized behavioral types called memory access protocols (\langs).
  %
  In this paper we extend the theory of \langs to
  characterize kernels for which alarms can be identified as true alarms.
  We introduce a core calculus for CUDA, which we named \babylang, and a
  behavioral type system for it.
  We show that if a \babylang program can be assigned a MAP, then any
  alarm raised by \Tool for this program is a true alarm.
\end{abstract}


\section{Introduction}
\label{sec:introduction}

%
%
%




\newcommand{\diaghorizsize}{\textwidth/2.1}
\newcommand{\diagvertsize}{4.1cm}
\newcommand{\exref}[1]{{\scriptsize Fig.~\ref{#1}}}
\newcommand{\exxref}[1]{{\scriptsize Eq.~\ref{#1}}}
\definecolor{racy}{rgb}{0.98, 0.93, 0.36}
\definecolor{frontier}{rgb}{0.55, 0.0, 0.55}
\definecolor{groundtruth}{rgb}{0.13, 0.55, 0.13}
%
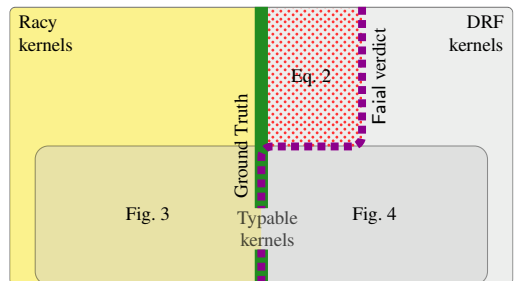
\begin{wrapfigure}[11]{r}{0.43\textwidth}
  \centering
  \begin{tikzpicture}[scale=0.9, every node/.style={transform shape},tips=proper,font=\scriptsize]
    \node (cuda) at (0,0) [draw=black,minimum width=\diaghorizsize,minimum
    height=\diagvertsize,venn=white,rounded corners=2pt] {};
    \node (racy) at (-\diaghorizsize/4,\diagvertsize/2) [minimum width=\diaghorizsize/2,minimum
    height=\diagvertsize,rounded corners=2pt,fill=racy,opacity=0.7,inner sep=4pt] {};
    \node (drf) at (\diaghorizsize/4,\diagvertsize/2) [minimum width=\diaghorizsize/2,minimum
    height=\diagvertsize,rounded corners=2pt,fill=green!10!black!10,opacity=0.7,inner sep=4pt] {};
    \node (inferable) at (0,0) [draw=black,minimum width=0.9\diaghorizsize,minimum
    height=\diagvertsize/2,venn=black!20,opacity=0.4,rounded corners=5pt] {};
    \node[anchor = north east, minimum
    height=\diagvertsize/2,yshift= \diagvertsize/2, xshift=\diaghorizsize/5 ,minimum width=\diaghorizsize/5,
    pattern=crosshatch dots,pattern color=red!80] at (0, \diagvertsize/2) {\exxref{eqn:illtyped}};
    %
    %
    %
    \node at ($(inferable.west)!0.25!(inferable.east)$) (exR) {\exref{lst:read-example-racy-infer-pre}};
    \node at ($(inferable.west)!0.75!(inferable.east)$) (exD) {\exref{lst:read-example-drf-infer-pre}};
    \node[inner sep=1pt,align=center,fill opacity=0.7,rounded corners=5pt,text opacity=1,yshift=-0.2cm]
    at ($(exR)!0.5!(exD)$) (typable) {Typable \\ kernels};
    \path[draw,line width=5pt,groundtruth] (cuda.south)  -- (typable.south);

    \path[draw,line width=5pt,groundtruth] (typable.north) -- (cuda.north);

    \node[rotate=90,yshift=0.3cm] at ($(cuda.north)!0.5!(cuda.south)$) (gt)  {Ground Truth};
    \path[draw,  line width=3pt, frontier, dotted, rounded corners]
    ($(cuda.north)!0.4!(cuda.north east)$) |- (0,\diagvertsize/2) -- (typable.north);
    \path[draw, line width=3pt, frontier, dotted] (typable.south) --  (cuda.south);
    \node[yshift=-\diagvertsize/4,xshift=0.25cm,rotate=90,  black!70] at ($ (cuda.north)!0.4!(cuda.north east) $ ) {{\color{black}\Tool verdict}};
    \node[below right,align=left] at (cuda.north west) {Racy \\ kernels};
    \node[below left,align=right] at (cuda.north east) {DRF \\ kernels};
  \end{tikzpicture}
  \caption{Completeness gap in \Tool~\cite{faial:cav21}.}
  \label{fig:venn-diagram}
\end{wrapfigure}


GPUs have become widespread in many computationally-demanding domains,
thanks to their parallel processing capabilities. They are notably
used in mission critical software (\eg self-driving cars); hence the
correctness of GPU programs is essential.
While several platforms (\eg CUDA) provide support for general
purpose programming using GPUs, writing correct GPU programs (aka.\
kernels) is notoriously error prone. This is notably due to their
unusual concurrency models and low-level programming language support.
Bugs in low-level concurrent programming are often due to data-races, \ie when
two or more threads access concurrently the same location in shared memory, and
at least one access is a write; causing an unexpected source of nondeterminism.
Such races can cause system failures that endanger both the safety and
security of critical systems.

Over the last decade, several authors proposed static approaches to verify
data-race freedom (DRF) of GPU
kernels~\cite{faial:cav21,gpuverify:toplas15,pug:fse10}.
While these techniques guarantee the absence of bugs, they may report false
alarms.
Further, research shows that false alarms hamper the adoption of static analysis
in industrial
settings~\cite{static-analysis-google:cacm18,why-static-analysis:icse13}.
The objective of this paper is to characterize a class of GPU
kernels that can be verified \emph{soundly} and
\emph{completely}.

Our approach is to extend the theory behind \Tool~\cite{faial:cav21}, a static
DRF analysis tool, so that alarms can be identified as true alarms, \ie provable
data-races.
\Tool centers its verification around an abstraction called \emph{memory access
protocols} (MAP): a form of behavioral types that expresses where data is
written to (resp.\ read from) shared arrays, but abstracts away the content of
arrays.
\Tool leverages the fact that \Lang  does not represent the array's data to
improve the performance of the DRF analysis.
The goal of this paper is to introduce a type system that identifies kernels for
which the DRF analysis in~\cite{faial:cav21} is sound and complete.

\miniparagraph{Contributions}
\cref{fig:venn-diagram} illustrates the contribution of this paper in
the context of static verification of GPU kernels.
It is generally undecidable to determine whether a given kernel is
racy or data-race-free (DRF).
For a given kernel, \Tool always gives a sound verdict (all bugs are
caught) but may report spurious data-races (\ie the dotted area in
\cref{fig:venn-diagram}).
Using a novel formal subset of CUDA, dubbed BabyCUDA, we characterize a
subset of kernels, for which \Tool's verdicts align with the ground
truth.
Our characterization takes the form of a syntax-directed typing system whereby
when a \babylang program is assigned a memory access protocol, then this protocol
reflects and preserves the behavior of its kernel.
We note that~\cite{faial:cav21} studies formally the verification of \langs, but
it only informally presents the inference step from CUDA programs to \langs.
This paper addresses that concern via \babylang, as our overarching
goal is to mechanize the entire \Tool tool-chain.


\section{CUDA Programming and Memory Access Protocols}
\label{sec:background}


%
CUDA follows the Single-Instruction-Multiple-Threads (SIMT)
programming model where multiple threads execute a copy of a GPU
program (the kernel) in lockstep.
%
%
Threads communicate over \emph{shared arrays} that all threads can
write to (resp.\ read from).
CUDA includes a variable that uniquely identifies each thread to help, among
other things, partition arrays among threads.
Threads can await each other via barrier synchronization with
\texttt{\_\_syncthreads()}.
A write to shared memory by one thread is only guaranteed to be visible by other
threads after a barrier synchronization.

\langs are behavioral types that codify how multiple threads interact over
shared memory.
\langs act as an over-approximation of the behavior of CUDA program.
In~\cite{faial:cav21} we have shown that it is possible to verify whether or not
\langs are DRF using a transformation to Satisfiability Modulo Theories (SMT).
%

%
\begin{figure}
  \centering
  \begin{ruled}
    \begin{tabular}[t]{llcl}
      $\NatDom \ni$ & $\Tid$ &$\grmeq$ & $0 \grmor 1 \grmor \cdots$
      \\
                    & $\Nexp$ & $\grmeq $& $\Var \grmor \Tid \grmor \Nexp \Nbin \Nexp$
      \\
                    &$\Mode$ & $\grmeq$ & $\Mwrite\grmor \Mread$
      \\
      $\Bool \ni$ & $\Bexp $&$\grmeq $& $\True \grmor \False \grmor \Nexp \Nrel \Nexp \grmor \Bexp \Bbin \Bexp$
      \\
      $\Clang \ni$ & $\Cexp$ &$\grmeq$ &
                                         $\Pskip \grmor
                                         \Acc{} \grmor
                                         \Cexp \Seq \Cexp
                                         \grmor
                                         \Cif[\Bexp]{\Cexp}{\Cexp} \grmor
                                         \Cfor[body=\Cexp]$\\
      $\Plang \ni$ & $\Proto$ &$\grmeq$ &
                                         $\Psync \grmor
                                          \Proto \Seq \Proto
                                          \grmor
                                          \Pfor[body={\Proto}]$
    \end{tabular}
    \caption{Syntax of \langs.}
    \label{fig:proto-syntax}
  \end{ruled}
\end{figure}


We review the syntax of \langs in~\cref{fig:proto-syntax}.
Hereafter, we use the following meta-variables for natural numbers
($\NatDom$, nonnegative integers): $\Nat$ for thread identifiers, $\Naty$ for array indices, and
$\Natz$ for array contents.
An arithmetic expression $\Nexp$ is either: a numeric variable~$\Var$, a thread
identifier variable~$\Tid$, or a binary operation on naturals yielding a
natural.
Boolean expression $\Bexp$ is either: a boolean literal, an
arithmetic comparison~$\Nrel$, or a propositional logic
connective~$\Bbin$.
We note that CUDA denotes the assignment and comparison operators as $=$ and $==$
respectively, whereas \langs and \babylang denotes the operators as $:=$ and $=$
respectively.

The language of protocols~\cite{faial:cav21} distinguishes syntactically between
protocol fragments that contain a barrier synchronization from those that do
not.
\Tool infers synchronized fragments automatically, and leverages such
distinction to greatly improve the performance of verifying DRF.
Unsynchronized protocols contain no synchronization constructs.
These include no-ops ($\Pskip$), array accesses,
unsynchronized sequencing, conditionals, and unsynchronized loops ($\KforU$).
Accessing an array is denoted as~$\Mode[\Naty]$ where $\Naty$ is an
index and $\Mode$ specifies whether the position is read ($\Mread$) or
written ($\Mwrite$).
Synchronized protocols $\Cexp \in \Clang$ on the other hand, must contain a
synchronization construct.
These include the barrier synchronization ($\Psync$), synchronized sequencing,
and sync-loops ($\KforS$).
Hereafter, we focus on inferring protocols for the unsynchronized fragment.
The treatment of the synchronized fragment follows similar (but simpler) lines.

\begin{figure}[t]
  \begin{ruled}
    \centering
    \begin{minipage}[t]{0.32\textwidth}
      \begin{code}[numbers=left]
for (int x=0; x<M; x++) {
   int y = A[x];
   A[x] = y + 1;
}
      \end{code}
    \end{minipage}
    \begin{minipage}[t]{0.32\textwidth}
      \begin{code}[numbers=left]
forU x inSymbol 0..M { ^\label{l:ex1-1}^
   let y = A[x] in     ^\label{l:ex1-2}^
   A[x] := y + 1       ^\label{l:ex1-3}^
}                     ^\label{l:ex1-4}^
      \end{code}
    \end{minipage}
    \begin{minipage}[t]{0.32\textwidth}
      \begin{code}[numbers=left]
forU x inSymbol 0..M {
   rd[x];
   wr[x]
}
      \end{code}
    \end{minipage}
    \vspace{-0.3cm}
    \caption{Racy example -- CUDA (left), BabyCUDA (middle), \lang (right)}
    \label{lst:read-example-racy-infer-pre}
    \label{lst:example-infer-1}
  \end{ruled}
\end{figure}


\cref{lst:read-example-racy-infer-pre} (left) illustrates a program
with a read and write access within a for-loop represented in CUDA.
This kernel is racy because all threads concurrently execute the same CUDA
program.
In this example, all threads read and write to array $\mathtt{A}$ at the
same index {\sf x} at each loop iteration.
In the protocol (right), only accesses (read/write and index) are
retained. Hence the race is also visible, and can be verified using an
off-the-shelf SMT solver.

\cref{lst:read-example-drf-infer-pre} (left) illustrates a minimal DRF
kernel using conditionals.
This kernel performs a write access on a single thread (with
identifier $\TidVar = 0$) while all other threads perform a no-op.
The kernel is DRF since no other array accesses occurs.

\section{\babylang: a Core Subset of CUDA}
\label{sec:syntax}

In this section we introduce the syntax and semantics of our calculus,
\babylang, that captures the concurrency mechanism and memory accesses
of CUDA.
%
%
%

\miniparagraph{Syntax}
We give the formal grammar of \babylang in~\cref{fig:b-syntax}.
Our analysis assigns one behavioral type per array, so
$\mathtt{A}$ denotes the only shared memory array accessible by all
threads.
We discuss the various constructs of the language with our two running
examples.

\begin{figure}[t]
  \begin{ruled}
    \centering
    \begin{minipage}[t]{0.32\textwidth}
      \begin{code}[numbers=left]
if (tid==0) {
  A[0] = tid;
}
      \end{code}
    \end{minipage}
    \begin{minipage}[t]{0.32\textwidth}
      \begin{code}[numbers=left]
if (tid=0) {     ^\label{l:ex2-1}^
  A[0] := tid    ^\label{l:ex2-2}^
} else { skip }  ^\label{l:ex2-3}^
      \end{code}
    \end{minipage}
    \begin{minipage}[t]{0.32\textwidth}
      \begin{code}[numbers=left]
if (tid=0)
 { wr[0] }
else
 { skip }
      \end{code}
    \end{minipage}
    \vspace{-0.3cm}
    \caption{DRF example -- CUDA (left), BabyCUDA (middle), \lang
      (right).}
    \label{lst:read-example-drf-infer-pre}
    \label{lst:example-infer-2}
  \end{ruled}
\end{figure}


\miniparagraph{\cref{lst:example-infer-1}}
\babylang loops have an upper and lower bounds, and a loop stride of~1.
In \cref{l:ex1-1}, the loop variable~$\Var$ starts at the lower bound of~$0$ and loops
until $\Var = M - 1$.
Besides loops, the only form of creating a local variable is by reading data
from an array.
In \cref{l:ex1-2}, we read from array~\texttt{A} and store the result in local
variable~$\Vary$.
%
%
%
In \cref{l:ex1-3}, we mutate the array by incrementing the value of element~$\Var$.




\miniparagraph{\cref{lst:example-infer-2}}
This example introduces \babylang conditionals and the $\Pskip$.
In \cref{l:ex2-1}, we show a conditional, which
unlike C, must always include the else branch.
Our calculus does not allow for mutation in the condition.
So, any reads/writes that would appear in a condition, must be hoisted before
the conditional.
The $\Pskip$ in \cref{l:ex1-2} denotes a no-op, as usual.
Sequencing---not seen in the examples---is standard.

%



\miniparagraph{Semantics}
We now discuss the semantics of \babylang.
We start by describing how we represent the runtime state of the array and formally define data-races.
We then introduce our operational semantics.
%

%
%
%
Evaluating a program yields an \emph{access history} (or, just
history), which is defined as a collection of memory accesses.
A memory access consists of the identifier of the thread,  whether it was a read
or a write, the index, and the data written if any.
Histories organize accesses in terms of synchronization phases, \ie a
history~$\Mhist$ is a list of phases.
We define a list inductively, as usual, the empty list is~$\Nil$, we construct
a list with~$\Hist\Cons\Mhist$, which prepends some element~$\Hist$ to a list~$\Mhist$.
We also use the usual notation $[\Hist_1,\dots,\Hist_n]$ to denote a list.
Each phase groups the accesses per thread, and also distinguishes reads from
writes.
A phase, ranged over by $\Hist$ and $\Histy$, maps each thread identifier~$\Tid
\in \TidDom$ into a pair of reads and writes $\RWset = (\Readset, \Writeset)$.
Pairs of reads-writes are ranged over by $\RWset$, $\RWsety$, and $\RWsetz$.
Reads are a set of naturals, ranged over by~$\Readset$, and denote the indices
read by a given thread.
Writes map naturals into naturals, are ranged over by~$\Writeset$, and denote
the indices and respective data written by a given thread.

The following is an example of a history.
Histories are ordered chronologically from right-to-left.
We present the earliest phase first.
Phase~0 states that thread~0 wrote number~9 to index 1,
and thread~1 wrote number 10 to index 2.
Phase~1 states that both threads wrote their thread identifier to index 0.
Additionally, thread~1 read from index~0.
\begin{equation}
    \label{eq:hist-example}
    [\overbrace{
    \{ 0\colon(\emptyset,\{ \WriteAcc{0}{0} \}) , 1\colon(\{0\},\{ \WriteAcc{0}{1} \}) \}
    }^{\text{phase 1}}
    ,
    \overbrace{
    \{ 0\colon(\emptyset,\{ \WriteAcc{1}{9} \}) , 1\colon(\emptyset,\{ \WriteAcc{2}{10} \}) \}
    }^{\text{phase 0}}
    ]
\end{equation}

For instance, in \cref{eq:hist-example}: phase 0 is \emph{not} racy;
phase~1 is racy (there is a write-write and a read-write data-races);
the history is not safe, as phase~1 is racy.

%
%

\begin{definition}[DRF program]
\label{def:drf}
$\Hist$ is \emph{racy} if
there exist identifiers $\Tid_1, \Tid_2 \in \TidDom$ and index $j \in \NatDom$ such that
$\Tid_1 \neq \Tid_2$,
$\Hist(\Tid_1) = \RW[r=\Readsetx,w=\Writesetx]$, $\Naty \in dom(\Writesetx)$,
$\Hist(\Tid_2) = \RW[r=\Readsety,w=\Writesety]$, and $\Naty \in \Readsety \cup dom(\Writesety)$.
%
%
%
$\Hist$ is DRF if $\Hist$ is not racy.
$\Mhist$ is DRF if every member of $\Mhist$ is DRF.
%
%
A program~$\BabyCexp$ is \emph{DRF} if $
        \PredIn[{[\emptyset]}] \BabyCexp \Pred
        \Mhist$ and
$\Mhist$ is DRF.
\end{definition}



%
%
The semantics, given in \cref{fig:b-semantics}, is divided into two big-step
operational semantics judgments
First, the evaluation of a single thread.
Second, the evaluation of a parallel program.

\miniparagraph{Single-threaded semantics}
The semantics of a single thread is given by~$\TredIn \BabyCexp \Tred \RWsety$,
where $\RWset$, $\BabyCexp$, $\Mhist$, and $\Tid$ are input parameters, and
$\RWsety$ is the result of evaluation.
The read-write pair~$\RWset$ represents the current phase.
Parameter~$\BabyCexp$ states the program being evaluated.
Parameter history~$\Mhist$ represents the phases preceding~$\RWset$, necessary
to look up the contents of the array, and remains unchanged throughout
evaluation.
The identifier~$\Tid$ states which thread is executing.
Rules for control-flow (sequencing, conditionals, and looping) and for skip are
standard.
Next, we explain the rules for reading from and write to an array.

Rule~\textsc{read} acts as a standard let-binder.
We evaluate the index expression~$\Nexp$ down to the index~$\Naty$,
which we use as the first parameter of function \textit{lastwrite}.
The second parameter of~\textit{lastwrite} is the current history:
we prepend the read-write set~$\RW$ of the current thread~$\Tid$ to the rest of the
history~$\Hist$, given as~$\{ \Tid \colon \RW \} \Cons \Mhist$.
In the continuation~$\BabyCexp$, we replace~$\Var$ by the last value written to
index~$\Naty$, and we also update the read set to include this access~$\Naty$.
%
%
The notation $\BabyCexp\subs{\Nexp}{\Var}$ stands for $\BabyCexp$ in which all
the free occurrences of $\Var$ are replaced by $\Nexp$.
%
%
The $\Lastwrite[\Naty]{\Mhist}$ function takes an index~$\Naty$ and a
history~$\Mhist$, and returns the last value written to~$\Naty$.
The function ranges over the phases of history~$\Mhist$, from most recent to
least recent.
When history~$\Mhist$ is DRF, it can be shown that there is at most
one writer per index, in every phase.
We denote the domain of a map as $dom(\Writeset)$.
Lastly, function~\textit{lastwrite} is partial, as it is undefined ($\bot$) when
no value has been written to index~$\Naty$.
Rule~\textsc{write} is simple, we evaluate the index~$\Nat$ down to~$\Naty$ and
the payload~$\Nexpy$ down to~$\Natz$ and then update the write map.
Map $\Writeset\update{\Naty}{\Natz}= \Writeset'$ denotes adding the pair~$\Naty$ and~$\Natz$ to
map~$\Writeset$, such that $\Writeset'(\Nat) = \Writeset(\Nat)$ if $\Nat \neq
\Naty$ otherwise $\Writeset'(\Nat) = \Natz$.
Rule~\textsc{seq} show phase continuity as it passes along the evaluation
chain, with $\RWsety$ being the intermediate phase and $\RWsetz$ the final
result.
Control-flow rules (\eg conditionals, loops, no-op) are standard.
%
%
We omit the synchronized \babylang fragment for presentation purposes.

\miniparagraph{Parallel semantics}
The judgment for parallel execution is~$\PredIn \BabyCexp\Pred\Mhisty$,
execute~$\BabyCexp$, given an input history~$\Mhist$, and produce an output
history~$\Mhisty$.
Rule~\textsc{par} runs~$\BabyCexp$ in each thread~$\Tid \in \TidDom$
independently, using the thread-semantics.
As is usual in this form of parallelism, we substitute a special
variable~$\TidVar$ by the thread identifier, $\BabyCexp\subs{\Tid}{\TidVar}$.
In order to hide writes by other threads in the current phase~$\Hist$, each
thread~$\Tid$ only has access to its own read-writes, $\Hist(\Tid)$.
All threads have access to the same remaining history~$\Mhist$.
Lastly, we ``merge'' the resulting read-writes of each thread to reconstruct the
current phase.
We note that our semantics ignores a form of memory movement caused by
data-races due to evaluating threads independently.
In CUDA, a thread \emph{may} indeed observe a racy-write from another thread.
However, our simplification still captures the occurrence of data-races, and as
such, does not affect the correctness of \drf analysis.

%

\begin{figure}
  \begin{ruled}
    \small
    \textbf{Syntax} \hfill
    \begin{minipage}[t]{0.5\textwidth}
      \begin{tabular}[t]{r@{\,}c@{\,}lll}
          $\BabyClang\ni\BabyCexp$ & $\grmeq$ & \
                                           $\Bwr[\Nexp]{\Nexp} \grmor$
                                           $\Brd[\Var]{\Nexp}{\BabyCexp}$ \\
                                   &$\grmor$ &\  $\BabyCexp \Seq \BabyCexp$
                                               $\grmor \Cif[\Bexp]{\BabyCexp}{\BabyCexp}$
        \\
                                   & $\grmor$&\ $\BabyCfor[body=\BabyCexp] \grmor \Pskip$ \\
        $\Mhist$ & $\grmeq$ & $\Nil  \grmor \Hist \Cons \Mhist$
      \end{tabular}
    \end{minipage}
    \begin{minipage}[t]{0.4\textwidth}
      \begin{tabular}[t]{lclll}
        $ \TidDom$ & $\subseteq$ & $\NatDom$\\
        $\Readset$ & $\in$ & $\mathcal{R} $ & $ \smash{\eqdef} $ & $ \mathcal{P}(\NatDom)$
        \\
        $\Writeset$ & $\in$ & $\mathcal{W} $ & $ \smash{\eqdef} $ & $ \NatDom \mapsto \NatDom$
        \\
        $\Hist$ & $\in$ & $ \TidDom $ & $ \mapsto $ & $ (\mathcal{R} \times \mathcal{W}) $
      \end{tabular}
    \end{minipage}

    \bigskip
    \bigskip

    \textbf{Semantics} \hfill \fbox{$\Lastwrite[\Naty]{\Mhist} = \Natz$}
    \quad
    \fbox{$\TredIn \BabyCexp \Tred \RWset$}
    \quad
    \fbox{ $\PredIn \BabyCexp \Pred \Mhist$ }
    \begin{mathpar}
      %
      \inferrule[lastwrite-curr]{
        \exists \Tid \st \Hist(\Tid) = \RW
        \qquad
        \Writeset(\Naty) = \Natz
      }
      {
        \Lastwrite[\Naty]{\Hist \Cons \Mhist} = \Natz
      }

      \inferrule[lastwrite-prev]{
        \forall \Tid \st \Hist(\Tid) = \RW
        \implies
        \Naty \notin dom(\Writeset)
      }
      {
        \Lastwrite[\Naty]{\Hist \Cons \Mhist} = \Lastwrite[\Naty]{\Mhist}
      }

      \inferrule[lastwrite-undef]{
        \,
      }
      {
        \Lastwrite[\Naty]{\Nil} = \bot
      }

      \inferrule[read]{
        \Nexp \Bstep \Naty
        \qquad
        \TredIn[{\RW[r={\Readset \cup \{ \Naty \}}]}]
        {\BabyCexp
          \subs
          {\Lastwrite[\Naty]{\{ \Tid \colon \RW \} \Cons \Mhist}}
          {\Var}
        } \Tred
        \RWsety
      }{
        \TredIn {\Brd{\Nexp}{\BabyCexp}} \Tred
        \RWsety
      }

      \inferrule[write]{
        \Nexp \Bstep \Naty
        \qquad
        \Nexpy \Bstep \Natz
      }{
        \TredIn[\RW] {\Bwr{\Nexpy}} \Tred
        \RW[w={\Writeset\update{\Naty}{\Natz}}]
      }

      \inferrule[seq]{
        \TredIn \BabyCexpx \Tred
        \RWsety
        \quad
        \TredIn[\RWsety] \BabyCexpy \Tred \RWsetz
      }{
        \TredIn {\BabyCexpx \Seq \BabyCexpy}  \Tred  \RWsetz
      }

      \inferrule[if-t]{
        \Bexp \Bstep \True
        \quad
        \TredIn  \BabyCexpx \Tred
        \RWsety
      }{
        \TredIn {\Cif{\BabyCexpx}{\BabyCexpy}}  \Tred  \RWsety
      }

      \inferrule[if-f]{
        \Bexp \Bstep \False
        \quad
        \TredIn \BabyCexpy \Tred
        \RWsety
      }{
        \TredIn {\Cif{\BabyCexpx}{\BabyCexpy}}  \Tred  \RWsety
      }

      \inferrule[for-1]{
        (\Nexp \ge \Nexpy) \Bstep \True
      }{
      \TredIn {\BabyCfor[body=\BabyCexp]} \Tred \RWset
      }

      \inferrule[for-2]{
        (\Nexp < \Nexpy)  \Bstep \True
        \qquad
        \TredIn{\BabyCexp\subs{\Nexp}{\Var}} \Tred \RWsety
        \qquad
        \TredIn[\RWsety]{\BabyCfor[lo=\Nexp +1,hi=\Nexpy,body=\BabyCexp]}
        \Tred \RWsetz
      }{
          \TredIn {\BabyCfor[body=\BabyCexp]}  \Tred  \RWsetz
      }

      \inferrule[skip]{
        \,
      }{
          \TredIn \Pskip  \Tred \RWset
      }

      %
      %
      %
      %
      %
      %
      \inferrule[par]{
        \Histy = \bigcup \{ \Tid \colon \RWset \, \mid \,
        \TredIn[\Hist(\Tid)] {\BabyCexp\subs{\Tid}{\TidVar}}
        \Tred \RWset \; \land\;
        \Tid \in \TidDom
        \}
      }{
        \PredIn[\Hist \Cons \Mhist] \BabyCexp \Pred
        \Histy \Cons \Mhist
      }
    \end{mathpar}

    %
    %
    %
    %
    %
    %
    %
    %
    %
    \caption{Syntax and semantics of \babylang (unsynchronized fragment).}
    \label{fig:b-syntax}
    \label{fig:b-semantics}
  \end{ruled}
\end{figure}





\section{A Behavioral Typing System for \babylang}
\label{sec:inference}

In this section, we introduce a type system that checks a \babylang program
against a behavioral type (\lang).
We also state our main result that our DRF analysis is sound and complete for
well-typed programs.

\begin{figure}
  \begin{ruled}
   \hfill
   \fbox{$\Ntypes \Nexp$}
   \quad
   \fbox{$\Btypes \Bexp$}
   \quad
   \fbox{$\Uinfer{\BabyCexp}  \InferTo \Cexp$}
    \hfill
    \begin{mathpar}
      \inferrule[t-n]{
        \FreeVar{\Nexp} \subseteq \Vset
      }{
        \Ntypes \Nexp
      }

      \inferrule[t-b]{
        \FreeVar{\Bexp} \subseteq \Vset
      }{
        \Btypes \Bexp
      }

      \inferrule[t-write]{
        \Ntypes \Nexp
      }{
        \Uinfer{\Bwr[\Nexp]{\Nexpy}} \InferTo
         \Mwrite[\Nexp]
      }

      \inferrule[t-read]{
        \Ntypes \Nexp
        \qquad
        \Vary \notin \Vset
        \qquad
        \Uinfer{\BabyCexp} \InferTo \Cexp
      }{
        \Uinfer{\Brd[\Vary]{\Nexp}{\BabyCexp}} \InferTo
         \Mread[\Nexp]  \Seq \Cexp
      }

      \inferrule[t-seq]{
        \Uinfer{\BabyCexpx}  \InferTo \Cexpx
        \quad
        \Uinfer{\BabyCexpy}  \InferTo \Cexpy
      }{
        \Uinfer{\BabyCexpx \Seq \BabyCexpy}  \InferTo  \Cexpx \Seq \Cexpy
      }

      \inferrule[t-if]{
        \Btypes \Bexp
        \quad
        \Uinfer{\BabyCexpx}  \InferTo \Cexpx
        \quad
        \Uinfer{\BabyCexpy}  \InferTo \Cexpy
      }{
        \Uinfer{\Cif{\BabyCexpx}{\BabyCexpy}} \InferTo  \Cif{\Cexpx}{\Cexpy}
      }

      \inferrule[t-for]{
        \Ntypes \Nexp
        \qquad
        \Ntypes \Nexpy
        \qquad
        \Var \notin \Vset
        \qquad
        \Uinfer[\Vset \cup \{\Var\}]{\BabyCexp} \InferTo  \Cexp
      }{
        \Uinfer{\BabyCfor[body=\BabyCexp]} \InferTo  \Cfor[body=\Cexp]
      }

      \inferrule[t-skip]{
        \,
      }{
        \Uinfer{\Pskip}  \InferTo \Pskip
      }

    \end{mathpar}

      %
      %

    %
    %
    %
    %
    %
    \caption{Behavioral type system for \babylang (unsynchronized fragment).}
    \label{fig:b-typecheck}
  \end{ruled}
\end{figure}

\cref{fig:b-typecheck} introduces our three typing judgments.
The idea behind our type system is to disallow indexing arrays using data read
from arrays --- data-dependent array indexing.
To this end, our type system maintains the set of variables allowed in
control flow and in indexing.
Let~$\Vset$ range over sets of variables.
A well-typed numeric expression $\Ntypes \Nexp$ only mentions variables defined
in~$\Vset$.
The $\FreeVar{\cdot}$ function returns the set of free variables of the
argument, defined for numeric and boolean expression.
Similarly, a well-typed boolean expression~$\Btypes \Bexp$ only mentions
variables defined in~$\Vset$.
Program~$\BabyCexp$ has a type~$\Cexp$ (an unsynchronized protocol), under an
environment~$\Vset$.
%
%
Rule~\textsc{t-write} constrains the use of the index expression~$\Nexp$.
The type of program~$\Bwr[\Nexp]{\Nexpy}$ is type~$\Mwrite[\Nexp]$, which does
not mention the payload~$\Nexpy$.
Rule~\textsc{t-read} also constrains the use of the index expression~$\Nexp$.
Since we do not extend the typing environment~$\Vset$ with~$\Var$, its
continuation~$\BabyCexp$ cannot use~$\Var$ in to affect control-flow and
indexing.
To simplify the formalism, we require nested binders are all distinct from each
other, hence~$\Var \notin \Vset$.
The type of~$\Brd[\Vary]{\Nexp}{\BabyCexp}$ sequences a read with the
type~$\Cexp$ of the continuation~$\BabyCexp$, that is, $\Mread[\Nexp] \Seq \Cexp$.
To ensure that nested binders are distinct, we have~$\Var \notin \Vset$.
The remaining rules yield a type that matches the shape of the program.
Rules~\textsc{t-seq} and~\textsc{t-if} are trivial: since there are no variables
being declared, we simply propagate the typing environment.
Rule~\textsc{t-for} constrains the upper and lower bound of the loop
and allows the use of variable~$\Var$ when typing the loop body~$\BabyCexp$.
Rule~\textsc{t-skip} states that $\Pskip$ is always well-typed.

\miniparagraph{Examples}
\cref{lst:example-infer-1} (middle) is well-typed under context~$\{M,\TidVar\}$.

\begin{equation*}
    \Uinfer[
    \{ M,\TidVar \}
    ]{
    \BabyCfor[
      lo=0,
      hi=M,
      body={
        \Brd[\Vary]{\Var}{
            \Bwr[\Var]{\Vary + 1}
        }
    }]}
    \InferTo
    {\Cfor[
      lo=0,
      hi=M,
      body={
        \Rd[\Var];
        \Wr[\Var]
        }
    ]}
\end{equation*}

\noindent
\cref{lst:example-infer-2} (middle) is well-typed under context~$\{\TidVar\}$.

\begin{equation*}
    \Uinfer[
      \{\TidVar\}
    ]{
      \Cif[\TidVar = 0]{\Bwr[0]{\TidVar}}{\Pskip}
    }
    \InferTo
    {
      \Cif[\TidVar = 0]{\Wr[0]}{\Pskip}
    }
\end{equation*}

\noindent
The following is an ill-typed program that falls within the dotted area
in~\cref{fig:venn-diagram}.
The kernel triggers a data-dependent false alarm in \Tool, caused by
variable~$\Var$ that was read from an array being used to index the array in
$\Bwr[\Var]{9}$.
In our experiments~\cite{faial:cav21}, we did not find any data-dependent false
alarms when dealing with real-world kernels.

\begin{equation}
    \label{eqn:illtyped}
        \NUinfer[\{\TidVar\}]{\Bwr[\TidVar]{\TidVar} \Seq \Brd[\Var]{\TidVar}{\Bwr[\Var]{9}}}
            \InferTo \Wr[\TidVar] \Seq \Rd[\TidVar] \Seq \Wr[\Var]
\end{equation}

%
Let $\Lambda$ range over a phase of \lang, defined as a set of access
values~$\AccVal \grmeq \By{\Acc[index=\Naty]}$, where $\Mode \in \{
\Mread,\Mwrite \}$.
Further, we define $\AccVal \mathbin{\hat\in}\Hist$ if $\AccVal =
\By{\Acc[index=\Naty]}$, $\Hist(\Tid) = \RW$, and either $o =\Mread \land
\Naty \in R$, or $o = \Mwrite \land \Naty \in dom(W)$.

\begin{theorem}[Correctness]
\label{thm:sound-type}
Let $\PredIn \BabyCexp \Pred \Mhist'$ and $\Cexp \Bstep \Lambda$.
If $\Sinfer[\{\TidVar\}]{\BabyCexp} \InferTo \Cexp$, $\Mhist$ is DRF,
then $\Mhist'$ is DRF if, and only if, $P$ is DRF.
\end{theorem}
\begin{sproof}
We prove a more general result from which this proof follows trivially.
Every access value in the source phase is in the target phase, and
vice-versa: if $\PredIn \BabyCexp \Pred {\Hist\Cons\Mhist'}$, $\Cexp \Bstep
\Lambda$, $\Sinfer[\{\TidVar\}]{\BabyCexp} \InferTo \Cexp$, and $\Mhist$ is
DRF, then $\AccVal \;\hat\in\; \Hist$ if, and only if, $\AccVal \in \Lambda$.
The proof follows by induction on the derivation of~$\PredIn \BabyCexp \Pred
\Hist\Cons\Mhist'$.

\end{sproof}

\miniparagraph{Discussion}
The main result establishes that well-typed programs are analyzed soundly and
completely.
The type system characterizes a subset of \babylang for which all data-race are
provably correct.
%
%
This theorem extends easily to the synchronized fragment.
%



\section{Related Work}
\label{sec:related-work}

\miniparagraph{Completeness in static analysis}
Gorogiannis \etal~\cite{gorogiannis:racerdx} introduce the first
DRF static analysis for multithreaded programs that is sound and complete, for a
subset of all programs.
We note, however, analyzing multithreading (where shared resources are protected
with locks) is generally inapplicable (or irrelevant) to GPU programming, and
vice-versa.
The focus of DRF analysis for multithreading is on lock usage, thread lifecycle,
and pointer aliasing, not on array access patterns.
Giacobazzi \etal~\cite{analyzing-analysis:popl15} develop a deductive system to
prove completeness of program analyses over an abstract domain.
Ranzato~\cite{ranzato:comp} uses completeness to better understand the
approximation of some intra-procedural analysis (\eg signedness, constant
propagation) applied to model checking.

\miniparagraph{DRF analysis for GPUs}
Static analysis include~\cite{gpuverify:toplas15,pug:fse10,faial:cav21}.
We note that \Tool exhibited the lowest rate of false alarms
in~\cite{faial:cav21}.
Symbol execution approaches can prove DRF but are unable to scale to larger
kernels~\cite{gklee:ppopp12,sesa:sc14,esbmc-gpu:sac16}.
Dynamic data-race detection, \eg \cite{%
  grace:ppopp11,%
  gmrace:tpds14,%
  ld:taco17,%
  curd:pldi18,%
  haccrg:icpp13,%
  barracuda:pldi17,%
  scord:isca20%
}, is sound and complete for all programs, but only for a single input and for a
single arbitrary thread schedule.
Finally, bug finding tools, such as \cite{simulee:icse20} sample thread
schedules to identify data-races, but are unable to prove DRF.

%


%

%
%

\section{Conclusion \& Future Work}
\label{sec:conclusion}



%
We tackle the problem of formally characterizing true data-races in the context
of static analysis for GPU kernels.
This paper introduces a core language (\babylang) and a behavioral
type system for \babylang.
Our main result is that our type system guarantees a sound and
complete DRF analysis. 

\miniparagraph{Future work}
Following the suggestions
of~\cite{static-analysis-google:cacm18,why-static-analysis:icse13}, we want to
improve upon the confidence of data-race reports, by identifying which accesses
can be analyzed soundly \emph{and} completely.
%
To this end, we will extend \Tool to annotate accesses according to the output
of the analysis introduced in this paper.
Additionally, we will continue our work in the Coq mechanization of our main
result.
%

%
%
%
%
%
%
%


\bibliographystyle{eptcs}
\bibliography{biblio}

\end{document}